\documentclass[conference,compsoc]{IEEEtran}
\IEEEoverridecommandlockouts

\ifCLASSOPTIONcompsoc
  \usepackage[nocompress]{cite}
\else
  \usepackage{cite}
\fi

\ifCLASSINFOpdf
\else
\fi

\usepackage[utf8]{inputenc}
\usepackage[T1]{fontenc}
\usepackage{amsmath,amssymb,amsfonts,mathtools,bm}
\usepackage{graphicx}
\usepackage{booktabs}
\usepackage{multirow}
\usepackage{colortbl}
\usepackage{float}
\usepackage{xcolor}
\usepackage{url}
\usepackage{microtype}
\usepackage{nicefrac}
\usepackage{stfloats}

\usepackage[numbers,square,sort&compress]{natbib}
\usepackage{hyperref}
\hypersetup{
    colorlinks=true,
    linkcolor=blue,
    citecolor=blue,
    urlcolor=blue,
}
\usepackage[capitalize,noabbrev]{cleveref}

\usepackage{caption}
\usepackage{subcaption}
\usepackage{algorithm}
\usepackage{algpseudocode}

\newcommand{\LineComment}[1]{\Statex \(\triangleright\) #1}
\newcommand{\LineCommentStep}[1]{\Statex \textbf{[Step #1]:}}

\usepackage{amsthm}
\theoremstyle{plain}

\theoremstyle{definition}

\theoremstyle{remark}

\usepackage{nccmath} 
\newenvironment{mbmatrix}{\begin{medsize}\begin{bmatrix}}{\end{bmatrix}\end{medsize}}

\newcommand{\mysplit}[1]{%
  \begin{tabular}{@{}c@{}}#1\end{tabular}
}

\begin{document}

\title{Volley Revolver: A Novel Matrix-Encoding Method  \\ for \\   Privacy-Preserving Deep Learning (Inference++)
}

\author{
\IEEEauthorblockN{John Chiang}
\IEEEauthorblockA{\textit{} \\
\textit{}\\
 \\
john.chiang.smith@gmail.com}
}

\maketitle

\begin{abstract}
Privacy-preserving inference of convolutional neural networks (CNNs) using homomorphic encryption has emerged as a promising approach for enabling secure machine learning in untrusted environments.
In our previous work~\cite{chiang2022volley}, we introduced a matrix-encoding strategy that allows convolution and matrix multiplication to be efficiently evaluated over encrypted data, enabling practical CNN inference without revealing either the input data or the model parameters.
The core idea behind this strategy is to construct a three-dimensional representation within ciphertexts that preserves the intrinsic spatial structure of both input image data and model weights, rather than flattening them into conventional two-dimensional encodings.
However, this approach can operate efficiently $only$ when the number of available plaintext slots within a ciphertext is sufficient to accommodate an entire input image, which becomes a critical bottleneck when processing high-resolution images.
In this paper, we address this fundamental limitation by proposing an improved encoding and computation framework that removes the requirement that a single encrypted ciphertext must fully contain one input image.
Our method reformulates the data layout and homomorphic operations to partition high-resolution inputs across multiple ciphertexts while preserving the algebraic structure required for efficient convolution and matrix multiplication.
Specifically, for the multiplication of two matrices $A \in \mathbb{R}^{m \times n}$ and $B \in \mathbb{R}^{n \times p}$, the proposed method uses $n$ ciphertexts to encode each matrix operand.
For a high-resolution image $I \in \mathbb{R}^{h \times w}$, our improved approach represents the encrypted image using $w$ ciphertexts, each encoding the corresponding column-wise image components with $h$ pixels packed across plaintext slots.
As a result, our approach enables privacy-preserving CNN inference to scale naturally beyond the slot-capacity constraints of prior methods, making homomorphic evaluation of CNNs practical for higher-resolution and more complex datasets.


\end{abstract}


%
\IEEEpeerreviewmaketitle

\section{Introduction}

\subsection{Background}

In application domains where sensitive information is routinely processed—such as healthcare, finance, and personalized services—machine learning systems must operate under strict privacy constraints while still delivering accurate predictions. This work focuses on privacy-preserving neural network inference, a setting in which a pre-trained model is deployed to a cloud platform to perform inference tasks on confidential user data. To ensure data confidentiality, the input data are encrypted prior to outsourcing, and the cloud server is assumed to be unable to access the plaintext information at any stage of computation. 

Among existing cryptographic techniques designed for secure computation, including Secure Multi-Party Computation (SMPC), Homomorphic Encryption (HE) offers a particularly strong security guarantee by enabling arbitrary computations to be carried out directly over encrypted data without decryption. This property makes HE especially well-suited for secure outsourced inference scenarios.

\subsection{Related Work}

The integration of Homomorphic Encryption with Convolutional Neural Network (CNN) inference has attracted significant research interest in recent years, beginning with the seminal work of Gilad-Bachrach et al.~\cite{gilad2016cryptonets}, who introduced the $\texttt{CryptoNets}$ framework. $\texttt{CryptoNets}$ demonstrated that neural networks can perform accurate inference on encrypted inputs while maintaining practical throughput. 

Subsequently, Chabanne et al.~\cite{chabanne2017privacy} extended this line of research to deeper CNN architectures by adopting the $\texttt{HElib}$ library~\cite{halevi2020helib}. Their work incorporated batch normalization and modified training procedures to obtain more stable and accurate polynomial approximations of the $\texttt{ReLU}$ activation function. Chou et al.~\cite{chou2018faster} further improved encrypted inference efficiency through pruning, quantization, and other deep learning optimization techniques, resulting in the $\texttt{Faster CryptoNets}$ framework. 

Brutzkus et al.~\cite{brutzkus2019low} proposed alternative data encoding strategies distinct from those used in $\texttt{CryptoNets}$ and introduced the Low-Latency CryptoNets ($\texttt{LoLa}$) approach to reduce inference latency. In a related direction, Jiang et al.~\cite{jiang2018secure} presented an efficient method for secure outsourced matrix multiplication based on a novel matrix encoding scheme, which significantly improves evaluation efficiency under homomorphic encryption. Chiang et al.~\cite{chiang2022volley} proposed a matrix-encoding strategy that enables efficient evaluation of convolution and matrix multiplication over encrypted data, thereby allowing practical CNN inference without revealing either the input data or the model parameters.

\subsection{Contributions}
This paper makes the following contributions:

\begin{itemize}
  \item We identify and formalize a fundamental scalability limitation of prior matrix-encoding approaches for homomorphic CNN inference, namely the requirement that a single ciphertext must contain an entire input image, which severely restricts applicability to high-resolution or large-scale inputs.

  \item We propose a novel encoding and computation framework that removes the single-ciphertext containment requirement by partitioning input data across multiple ciphertexts while preserving the algebraic structure necessary for efficient homomorphic convolution and matrix multiplication.

  \item We introduce a multi-ciphertext formulation for matrix multiplication, in which each operand matrix is encoded using $n$ ciphertexts for matrices in $\mathbb{R}^{m \times n}$ and $\mathbb{R}^{n \times p}$, enabling scalable homomorphic evaluation without sacrificing correctness.

  \item We design an efficient column-wise ciphertext representation for high-resolution images, where an image $I \in \mathbb{R}^{h \times w}$ is encrypted using $w$ ciphertexts, each packing $h$ pixels across plaintext slots, allowing privacy-preserving CNN inference to scale naturally with image resolution.

  \item We demonstrate that the proposed framework maintains the correctness and composability of homomorphic CNN operations, while substantially improving scalability beyond the plaintext slot-capacity constraints of existing methods.
\end{itemize}

\section{Preliminaries}

Throughout this paper, the symbols ``$\oplus$'' and ``$\otimes$'' are used to denote element-wise addition and element-wise multiplication, respectively, between ciphertexts encrypting matrix-valued plaintexts. We write $\texttt{ct}.P$ to represent the ciphertext corresponding to an encrypted matrix $P$. Furthermore, $I_{[i][j]}^{(m)}$ denotes the pixel located at column $j$ and row $i$ of the $m$-th image in the dataset.

\subsection{Fully Homomorphic Encryption}

Homomorphic Encryption (HE) is a cryptographic primitive that enables arithmetic operations to be performed directly on encrypted data. Under an HE scheme, computations carried out over ciphertexts yield new ciphertexts that decrypt to the correct results of the corresponding plaintext operations, without requiring decryption during computation or access to the secret key. Since Gentry~\cite{gentry2009fully} introduced the first fully homomorphic encryption scheme, thereby resolving a problem that had remained open for more than three decades, significant progress has been achieved in developing efficient encoding techniques for applying machine learning models under HE.

Cheon et al.~\cite{ckks2017homomorphic} proposed the CKKS scheme, which efficiently supports approximate arithmetic on encrypted real numbers. A key component of this scheme is the introduction of a procedure known as \texttt{rescaling}, which controls the growth of plaintext magnitudes during homomorphic evaluation. Their open-source implementation, $\texttt{HEAAN}$, similarly to other modern HE libraries, supports the Single Instruction Multiple Data (SIMD) paradigm~\cite{SmartandVercauteren_SIMD}, enabling multiple values to be packed into and processed within a single ciphertext.

Given a security parameter, $\texttt{HEAAN}$ generates a secret key $\textit{sk}$, a public key $\textit{pk}$, and additional public evaluation keys for operations such as ciphertext rotation. For ease of presentation, we omit explicit discussion of the \texttt{rescale} operation and assume that plaintext magnitude management is handled automatically. The core homomorphic functionalities provided by $\texttt{HEAAN}$ are summarized as follows:
\begin{enumerate}
    \item \texttt{Enc}$_\textit{pk}(m)$: Encrypts a plaintext message vector $m$ using the public key $\textit{pk}$ and outputs a ciphertext $\texttt{ct}$.
    
    \item \texttt{Dec}$_\textit{sk}(\texttt{ct})$: Decrypts a ciphertext $\texttt{ct}$ with the secret key $\textit{sk}$ and returns the corresponding plaintext vector.
    
    \item \texttt{Add}($\texttt{ct}_1, \texttt{ct}_2$): Produces a ciphertext encrypting the element-wise sum $\texttt{Dec}_\textit{sk}(\texttt{ct}_1) \oplus \texttt{Dec}_\textit{sk}(\texttt{ct}_2)$.
    
    \item \texttt{Mul}($\texttt{ct}_1, \texttt{ct}_2$): Outputs a ciphertext encrypting the component-wise product $\texttt{Dec}_\textit{sk}(\texttt{ct}_1) \otimes \texttt{Dec}_\textit{sk}(\texttt{ct}_2)$.
    
    \item \texttt{cMul}($C, \texttt{ct}_2$): Returns a ciphertext encrypting the element-wise multiplication between a plaintext constant matrix $C$ and the decrypted message of $\texttt{ct}_2$.
    
    \item \texttt{Rot}($\texttt{ct}, l$): Generates a ciphertext encrypting a plaintext vector obtained by cyclically rotating the message encrypted in $\texttt{ct}$ to the left by $l$ slots.
\end{enumerate}

\subsection{Database Encoding Method}

For clarity, we assume a training dataset consisting of $n$ samples, each represented by $f$ features, and that the number of available slots in a single ciphertext is at least $n \times f$. Such a dataset can be naturally organized as a matrix $Z \in \mathbb{R}^{n \times f}$, where each row corresponds to one training example. Kim et al.~\cite{IDASH2018Andrey} proposed an efficient database encoding strategy that packs the matrix $Z$ into a single ciphertext in a row-wise manner.

Their method introduces two fundamental shifting operations achieved through ciphertext rotations by $1$ and $f$ slots, respectively, namely incomplete column shifting and row shifting. The effect of applying the incomplete column shifting operation to matrix $Z$ is illustrated below:
 
\begin{align*}
 Z &= 
 \begin{bmatrix}
 z_{[1][1]}    &   z_{[1][2]}   &  \ldots  & z_{[1][f]}   \\
 z_{[2][1]}    &   z_{[2][2]}   &  \ldots  & z_{[2][f]}   \\
 \vdots    &   \vdots   &  \ddots  & \vdots   \\
 z_{[n][1]}    &   z_{[n][2]}   &  \ldots  & z_{[n][f]}   \\
 \end{bmatrix}   \\
 &\xmapsto{\textsl{incomplete} \text{ column shifting} }
 \begin{bmatrix}
 z_{[1][2]}    &   z_{[1][3]}   &  \ldots  & z_{[2][1]}   \\
 z_{[2][2]}    &   z_{[2][3]}   &  \ldots  & z_{[3][1]}   \\
 \vdots    &   \vdots   &  \ddots  & \vdots   \\
 z_{[n][2]}    &   z_{[n][3]}   &  \ldots  & z_{[1][1]}   \\
 \end{bmatrix} .  
\end{align*} 

Han et al.~\cite{han2018efficient} further summarized two aggregation procedures, namely $\texttt{SumRowVec}$ and $\texttt{SumColVec}$, which compute row-wise and column-wise summations, respectively. Applying these procedures to $Z$ yields the following results:
\begin{align*}
\texttt{SumRowVec(${Z}$)} &= \\
&\hspace{-1.2cm}
\begin{bmatrix}
 \sum_{i=1}^n z_{[i][1]}    &   \sum_{i=1}^n z_{[i][2]}   &  \ldots  & \sum_{i=1}^n z_{[i][f]}   \\
 \sum_{i=1}^n z_{[i][1]}    &   \sum_{i=1}^n z_{[i][2]}   &  \ldots  & \sum_{i=1}^n z_{[i][f]}   \\
 \vdots    &   \vdots   &  \ddots  & \vdots   \\
 \sum_{i=1}^n z_{[i][1]}    &   \sum_{i=1}^n z_{[i][2]}   &  \ldots  & \sum_{i=1}^n z_{[i][f]}   \\
\end{bmatrix}, \\[1ex]
\texttt{SumColVec(${Z}$)} &= \\
&\hspace{-1.2cm}
\begin{bmatrix}
 \sum_{j=1}^f z_{[1][j]}    &   \sum_{j=1}^f z_{[1][j]}    &  \ldots  & \sum_{j=1}^f z_{[1][j]}    \\
 \sum_{j=1}^f z_{[2][j]}    &   \sum_{j=1}^f z_{[2][j]}   &  \ldots  & \sum_{j=1}^f z_{[2][j]}   \\
 \vdots    &   \vdots   &  \ddots  & \vdots   \\
 \sum_{j=1}^f z_{[n][j]}    &   \sum_{j=1}^f z_{[n][j]}   &  \ldots  & \sum_{j=1}^f z_{[n][j]}   \\
\end{bmatrix}.
\end{align*}

To further support homomorphic convolution, we introduce a new aggregation primitive, denoted as $\texttt{SumForConv}$, which is designed to facilitate convolution operations over encrypted images, as detailed in Algorithm~\ref{alg:SumForConv}. The computational overhead of $\texttt{SumForConv}$ consists of approximately $2k$ homomorphic additions (\texttt{Add}), one plaintext–ciphertext multiplication (\texttt{cMul}), and $2k$ ciphertext rotations (\texttt{Rot}). Since rotations incur substantially higher cost than additions and multiplications, the overall complexity is dominated by $O(k)$ rotation operations.

As an illustrative example, consider the case where $n=f=4$ and the kernel size is $3 \times 3$. Applying $\texttt{SumForConv}$ to $Z$ produces:
\begin{align*}
 Z &= 
 \begin{bmatrix}
 z_{[1][1]}    &   z_{[1][2]}   &  z_{[1][3]}  & z_{[1][4]}   \\
 z_{[2][1]}    &   z_{[2][2]}   &  z_{[2][3]}  & z_{[2][4]}   \\
 z_{[3][1]}    &   z_{[3][2]}   &  z_{[3][3]}  & z_{[3][4]}   \\
 z_{[4][1]}    &   z_{[4][2]}   &  z_{[4][3]}  & z_{[4][4]}   \\
 \end{bmatrix} \\
 &\xmapsto{ \texttt{SumForConv}(\cdot,3,3) }
 \begin{bmatrix}
 s_{[1][1]}    &   s_{[1][2]}   &  0       & 0   \\
 s_{[2][1]}    &   s_{[2][2]}   &  0       & 0   \\
 0         &   0        &  0       & 0   \\
 0         &   0        &  0       & 0   \\
 \end{bmatrix},
\end{align*} 
where $s_{[i][j]} = \sum_{p=i}^{i+2} \sum_{q=j}^{j+2} z_{[p][q]}$ for $1 \le i,j \le 2$. Within a convolutional layer, $\texttt{SumForConv}$ enables the simultaneous computation of partial convolution results for an encrypted image.
 
\begin{algorithm}[htp]
    \caption{SumForConv: sum some part results of convolution operation after one element-wise multiplication}
     \begin{algorithmic}[1]
        \Require a ciphertect $\texttt{ct}.I$ encrypting a (convolved)  image $I$ of size $h \times w$,   the size $k \times k$ of some kernal $K$ with its bias $k_0$, and a stride of $(1, 1)$
        \Ensure a ciphertext $\texttt{ct}.I_s$ encrypting a resulting  image $I_s$ of the same size as $I$

        \State Set $ I_s \gets \boldsymbol 0 $
        \Comment{$I_s \in \mathbb R^{(h-k+1) \times (w-k+1)}$}
        \For{$i := 1$ to $(h-k+1)$}
           \For{$j := 1$ to $(w-k+1)$}
              \State $ I_s[i][j] \gets k_0 $
           \EndFor
        \EndFor
        \State $\texttt{ct}.I_s  \gets  \texttt{Enc}_\textit{pk}(I_s) $
        \LineComment Accumulate columns (could be computed in parallel)
        \For{$pos := 0$ to $k-1$}
           \State $\texttt{ct}.T \gets \texttt{Rot}(\texttt{ct}.I, pos)$  
           \State $\texttt{ct}.I_s  \gets \texttt{Add}(\texttt{ct}.I_s, \texttt{ct}.T)$
        \EndFor
        
        \LineComment Accumulate rows (could be computed in parallel)
        \For{$pos := 1$ to $k-1$}
           \State $\texttt{ct}.T \gets \texttt{Rot}(\texttt{ct}.I, pos \times w)$
           \State $\texttt{ct}.I_s  \gets \texttt{Add}(\texttt{ct}.I_s, \texttt{ct}.T)$
        \EndFor
        
        \LineComment Build a new designed matrix to filter out the garbage values
        \State Set $ M \gets \boldsymbol 0 $
        \Comment{$M \in \mathbb R^{h \times w }$}
        \For{$hth := 0$ to $(h-1)$}
           \For{$wth := 0$ to $(w-1)$}
              \If{$wth \bmod k = 0 $ {\bf and}  $wth + k \leq width$  {\bf and} $hth \bmod k = 0$   {\bf and} $hth + k \leq height $}
                 \State $ M[hth][wth] \gets  1 $
              \EndIf 
           \EndFor
        \EndFor
        \State $\texttt{ct}.I_s  \gets  \texttt{cMul}(M, \texttt{ct}.I_s)$
 
        \State \Return $ \texttt{ct}.I_s $
        \end{algorithmic}
       \label{alg:SumForConv}
\end{algorithm}

\subsection{Convolutional Neural Networks }
Convolutional Neural Networks (CNNs) are a specialized class of neural models designed primarily for image analysis tasks. In addition to the two fundamental layer types---Fully Connected (FC) layers and Activation (ACT) layers---CNNs incorporate two domain-specific components: Convolutional (CONV) layers and Pooling (POOL) layers. A typical CNN architecture for image classification can be expressed as
$[[CONV \to ACT]^p \to POOL]^q \to [CONV \to ACT] \to [FC \to ACT]^r \to FC$
where $p$, $q$, and $r$ are integers usually greater than one. In our implementation, we follow the network configuration adopted in~\cite{jiang2018secure}, which simplifies the architecture to
$[CONV \to ACT] \to [FC \to ACT] \to FC.$

The convolutional layer serves as the core computational building block of a CNN. It is characterized by kernels of spatial size $k \times k$, a stride parameter $(s, s)$, and a channel (or map) count $c$. Each kernel contains $k \times k \times c$ trainable weights together with a bias term $k_0$, all of which are optimized during training.

Given a grayscale input image $I \in \mathbb{R}^{h \times w}$ and a convolution kernel $K \in \mathbb{R}^{k \times k}$, applying convolution with a unit stride yields an output image $I' \in \mathbb{R}^{h' \times w'}$, where
\[
I'_{[i'][j']} = k_0 + \sum_{i=1}^{k} \sum_{j=1}^{k}
K_{[i][j]} \cdot I_{[i' + i][j' + j]},
\]
for $0 < i' \le h - k + 1$ and $0 < j' \le w - k + 1$. This formulation naturally extends to color images and multi-channel feature maps. When multiple kernels are used, the convolutional layer aggregates their respective feature maps by stacking the outputs.

Since FC layers operate exclusively on one-dimensional vectors, the multidimensional output produced by preceding layers (such as $[[\text{CONV} \rightarrow \text{ACT}]^p \rightarrow \text{POOL}]$ or $[\text{CONV} \rightarrow \text{ACT}]$) must be flattened before being passed to an FC layer.

\section{Technical Details}

We propose a novel matrix encoding scheme, referred to as \texttt{Volley Revolver}, which is particularly well suited for secure matrix multiplication under homomorphic encryption. The key idea is to arrange each semantically self-contained unit of information---for instance, an individual data sample---into a dedicated row of a matrix, which is then encrypted into a single ciphertext.

When applied to privacy-preserving neural network inference, \texttt{Volley Revolver} encodes the complete set of weights associated with each neuron into the corresponding matrix row. All neurons belonging to the same network layer are thus organized into a single matrix and encrypted collectively as one ciphertext.

\subsection{Encoding Method for Matrix Multiplication}
\label{section:EncodingMethodforMatrixMultiplication}

Consider an $m \times n$ matrix $A$ and an $n \times p$ matrix $B$. Their matrix product $C = A \cdot B$ is an $m \times p$ matrix with entries
\[
C_{[i][j]} = \sum_{k=1}^{n} a_{[i][k]} \cdot b_{[k][j]}.
\]

The matrices are explicitly defined as
\begin{align*}
  A &= 
 \begin{bmatrix}
 a_{[1][1]}    &   a_{[1][2]}   &  \ldots  & a_{[1][n]}   \\
 a_{[2][1]}    &   a_{[2][2]}   &  \ldots  & a_{[2][n]}   \\
 \vdots    &   \vdots   &  \ddots  & \vdots   \\
 a_{[m][1]}    &   a_{[n][2]}   &  \ldots  & a_{[m][n]}   \\
 \end{bmatrix} 
 ,  \\
 B &= 
 \begin{bmatrix}
 b_{[1][1]}    &   b_{[1][2]}   &  \ldots  & b_{[1][p]}   \\
 b_{[2][1]}    &   b_{[2][2]}   &  \ldots  & b_{[2][p]}   \\
 \vdots    &   \vdots   &  \ddots  & \vdots   \\
 b_{[n][1]}    &   b_{[n][2]}   &  \ldots  & b_{[n][p]}   \\
 \end{bmatrix}.
\end{align*}   

For ease of exposition, we assume that matrices $A$, $B$, and $C$ can each be encrypted into a single ciphertext, and that $m > p$. The case where $m \le p$ can be handled analogously and is therefore omitted.

In the homomorphic setting, \texttt{Volley Revolver} directly encodes matrix $A$, while matrix $B$ is first transposed, padded, and then encoded using a distinct row-ordering strategy. For matrix $A$, we adopt the encoding map proposed in~\cite{jiang2018secure}, defined as
\[
\tau_a : A \mapsto \bar{A} =
\left(
a_{[1 + \lfloor k/n \rfloor][1 + (k \bmod n)]}
\right)_{0 \le k < m \times n}.
\]

For matrix $B$, however, \texttt{Volley Revolver} employs a substantially different encoding approach. Specifically, $B$ is transposed and vertically tiled to construct an $m \times n$ matrix. The corresponding encoding map is given by
\[
\tau_b : B \mapsto \bar{B} =
\left(
b_{[1 + (k \bmod n)][1 + (\lfloor k/n \rfloor \bmod p)]}
\right)_{0 \le k < m \times n}.
\]

This mapping produces the following structure:
\begin{align*}
 B &= 
 \begin{bmatrix}
 b_{[1][1]}    &   b_{[1][2]}   &  \ldots  & b_{[1][p]}   \\
 b_{[2][1]}    &   b_{[2][2]}   &  \ldots  & b_{[2][p]}   \\
 \vdots    &   \vdots   &  \ddots  & \vdots   \\
 b_{[n][1]}    &   b_{[n][2]}   &  \ldots  & b_{[n][p]}   \\
 \end{bmatrix}  
 \xmapsto{\tau_b }  \\
 &\begin{bmatrix}
 b_{[1][1]}    &   b_{[2][1]}   &  \ldots  & b_{[n][1]}   \\
 b_{[1][2]}    &   b_{[2][2]}   &  \ldots  & b_{[n][2]}   \\
 \vdots    &   \vdots   &  \ddots  & \vdots   \\
 b_{[1][p]}    &   b_{[2][p]}   &  \ldots  & b_{[n][p]}   \\
 b_{[1][1]}    &   b_{[2][1]}   &  \ldots  & b_{[n][1]}   \\
 \vdots    &   \vdots   &  \ddots  & \vdots   \\
 b_{[1][1 + (m-1)\%p]}    &   b_{[2][1 + (m-1)\%p]}   &  \ldots  & b_{[n][1 + (m-1)\%p]}   \\
 \end{bmatrix}.
\end{align*}

\subsection{Homomorphic Matrix Multiplication}
We present an efficient evaluation algorithm for matrix multiplication in the encrypted domain. The algorithm maintains an accumulator ciphertext, denoted $\texttt{ct}.R$, which initially encrypts either zeros or a predefined constant value, such as the bias term in a fully connected layer. In addition, we introduce an operation termed \texttt{RowShifter}, which performs a specialized row-wise permutation on the encrypted matrix $\bar{B}$.

Specifically, \texttt{RowShifter} removes the first row of $\bar{B}$ and appends an appropriate existing row from the same matrix to the bottom. The transformation is illustrated as follows:
\begin{align*}
 {\bar{B}} &= 
 \begin{bmatrix}
 b_{[1][1]}    &   b_{[2][1]}   &  \ldots  & b_{[n][1]}   \\
 b_{[1][2]}    &   b_{[2][2]}   &  \ldots  & b_{[n][2]}   \\
 \vdots    &   \vdots   &  \ddots  & \vdots   \\
 b_{[1][p]}    &   b_{[2][p]}   &  \ldots  & b_{[n][p]}   \\
 b_{[1][1]}    &   b_{[2][1]}   &  \ldots  & b_{[n][1]}   \\
 \vdots    &   \vdots   &  \ddots  & \vdots   \\
 b_{[1][r]}    &   b_{[2][r]}   &  \ldots  & b_{[n][r]}   \\
 \end{bmatrix}
 \xmapsto{\texttt{RowShifter}( {\bar B} )}  \\
 &\begin{bmatrix}
 b_{[1][2]}    &   b_{[2][2]}   &  \ldots  & b_{[n][2]}   \\
 \vdots    &   \vdots   &  \ddots  & \vdots   \\
 b_{[1][p]}    &   b_{[2][p]}   &  \ldots  & b_{[n][p]}   \\
 b_{[1][1]}    &   b_{[2][1]}   &  \ldots  & b_{[n][1]}   \\
 \vdots    &   \vdots   &  \ddots  & \vdots   \\
 b_{[1][r]}    &   b_{[2][r]}   &  \ldots  & b_{[n][r]}   \\
 b_{[1][(r+1)\%p]}    &   b_{[2][(r+1)\%p]}   &  \ldots  & b_{[n][(r+1)\%p]}   \\
 \end{bmatrix} .
\end{align*} 

Algorithm~\ref{alg:RowShifter} details the procedure by which \texttt{RowShifter} constructs a new ciphertext from $\texttt{ct}.\bar{B}$.

\begin{algorithm}[htp]
    \caption{RowShifter: To shift row like a revolver}
     \begin{algorithmic}[1]
        \Require a ciphertext $\texttt{ct}.M$ encrypting a  matrix $M$ of size $m \times n$,   the number $p$, 
        and the number $idx$ that is determined in the Algorithm \ref{alg:Homomorphic matrix multiplication}
        \Ensure a ciphertext $\texttt{ct}.R$ encrypting the resulting  matrix $R$ of the same size as $M$

        \State Set $ R \gets \boldsymbol 0 $
        \Comment{$R \in \mathbb R^{m \times n}$}
        \State $\texttt{ct}.R  \gets  \texttt{Enc}_\textit{pk}(R) $
        
        \LineCommentStep1 Rotate the ciphertext first and then filter out the last row
        \State $\texttt{ct}.T \gets \texttt{Rot}(\texttt{ct}.M, n)$
        \LineComment Build a specially designed matrix to filter out the last row
        \State Set $ F_1 \gets \boldsymbol 1 $
        \Comment{$F_1 \in \mathbb R^{m \times n}$}
        \For{$j := 1$ to $n$}
            \State $ F_1[m][j] \gets 0 $
        \EndFor
        \State $\texttt{ct}.T_1  \gets  \texttt{cMul}(F_1, \texttt{ct}.T)$
          
        \LineCommentStep2 Rotate the ciphertext first and then filter out the last row
        \State $\texttt{ct}.P \gets \texttt{Rot}(\texttt{ct}.M, n \times ((m\%p + idx +1)\% p - idx)$  
        \LineComment Build a specially designed matrix to filter out the last row
        \State Set $ F_2 \gets \boldsymbol 0 $
        \Comment{$F_2 \in \mathbb R^{m \times n}$}
        \For{$j := 1$ to $n$}
            \State $ F_2[m][j] \gets 1 $
        \EndFor
        \State $\texttt{ct}.T_2  \gets  \texttt{cMul}(F_2, \texttt{ct}.P)$
          
        \LineComment  concatenate            
        \State $\texttt{ct}.R  \gets  \texttt{Add}(\texttt{ct}.T_1, \texttt{ct}.T_2)$
 
        \State \Return $ \texttt{ct}.R $
        \end{algorithmic}
       \label{alg:RowShifter}
\end{algorithm}
 
Given two ciphertexts, denoted as $\texttt{ct}.A$ and $\texttt{ct}.\bar{B}$, the proposed homomorphic matrix multiplication procedure is executed over $p$ consecutive iterations. For each iteration index $k$ satisfying $0 \leq k < p$, the algorithm performs the following four stages.

\noindent\texttt{Step 1: Row-wise alignment and multiplication.}  
In this stage, the operator $\texttt{RowShifter}$ is applied to the ciphertext $\texttt{ct}.\bar{B}$, producing a shifted ciphertext $\texttt{ct}.\bar{B}_1$. Subsequently, a homomorphic multiplication is carried out between $\texttt{ct}.A$ and $\texttt{ct}.\bar{B}_1$, resulting in the ciphertext $\texttt{ct}.A\bar{B}_1$. When $k=0$, the shifting operation degenerates to an identity transformation, and $\texttt{RowShifter}$ simply outputs a copy of $\texttt{ct}.\bar{B}$.

\noindent\texttt{Step 2: Regional summation.}  
The cloud server then applies the $\texttt{SumColVec}$ procedure to the ciphertext $\texttt{ct}.A\bar{B}_1$. This operation computes partial sums over regions determined by the kernel size of the convolution, generating an intermediate ciphertext denoted as $\texttt{ct}.D$.

\noindent\texttt{Step 3: Redundancy filtering.}  
To eliminate redundant elements within $\texttt{ct}.D$, a specially constructed filtering matrix $F$ is introduced. A single constant multiplication $\texttt{cMul}(F, \texttt{ct}.D)$ is performed, yielding a refined ciphertext $\texttt{ct}.D_1$.

\noindent\texttt{Step 4: Accumulation.}  
Finally, the ciphertext $\texttt{ct}.D_1$ is accumulated into the running result ciphertext $\texttt{ct}.R$, which aggregates the partial outputs obtained across iterations.

The algorithm iteratively executes Steps~1 through~4 for a total of $p$ rounds. After all intermediate results have been accumulated, the final output ciphertext $\texttt{ct}.C$ is produced. The complete procedure is formally described in Algorithm~\ref{alg:Homomorphic matrix multiplication}.

Table~\ref{tab2} summarizes the computational complexity and multiplicative depth required by each step of Algorithm~\ref{alg:Homomorphic matrix multiplication}.

\begin{table}[bht]
\centering
\caption{ Complexity and required depth of Algorithm~\ref{alg:Homomorphic matrix multiplication} vs Our Improved Version }
\label{tab2}
\begin{tabular}{|c||c|c|c|c|}
\hline
$\texttt{Step}$     & $\texttt{Add}$  & $\texttt{cMult}$  & $\texttt{Rot}$  & $\texttt{Mult}$   \\
\hline\hline
$\texttt{1}$        &   1       &  2       &   2       &   1        \\
\hline
$\texttt{2}$        &    2 $log_2 p$    &   1       &  2$ \log_2 p$      & 0   \\
\hline
$\texttt{3}$        &  0   &  1       &  0       &  0             \\
\hline
$\texttt{4}$        &  1       &  0       &   0       &  0           \\
\hline\hline
$\texttt{Total}$    &   $O(p\log p)$    &   $O(p)$    &   $O(p\log p)$       &    $O(p)$              \\
\hline
\end{tabular}
\end{table}

\begin{figure*}[htp]
\centering
\includegraphics[scale=0.4]{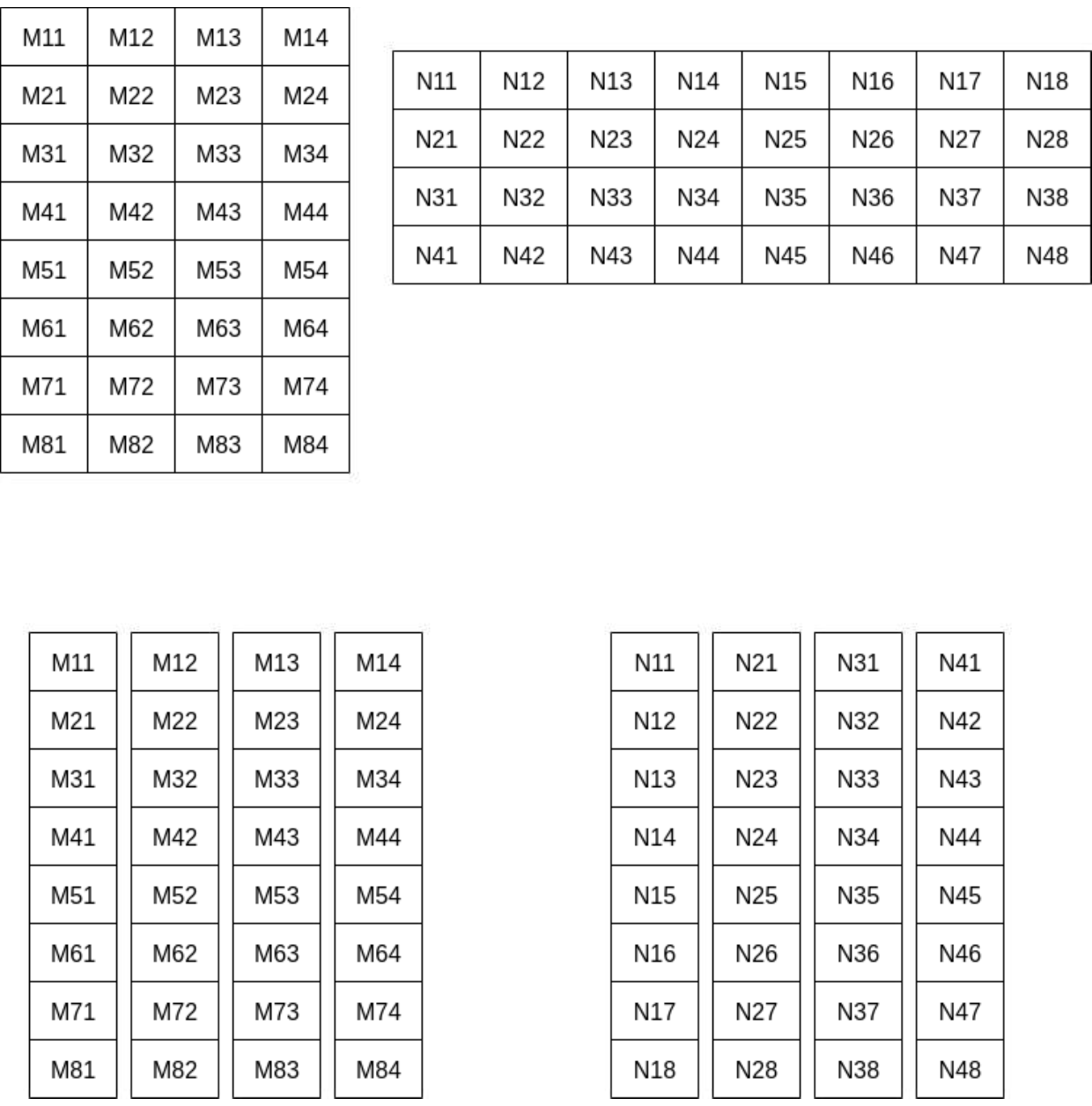}
\caption{
 Our matrix multiplication algorithm with $m = 2$, $n = 4$ and $p = 2$}
\label{Matrix Multiplication}
\end{figure*}

Figure~\ref{Matrix Multiplication} illustrates a representative example of Algorithm~\ref{alg:Homomorphic matrix multiplication} for the case where $m=2$, $n=4$, and $p=2$.

\begin{algorithm}[tbh]
    \caption{Homomorphic matrix multiplication}
     \begin{algorithmic}[1]
        \Require  $\texttt{ct}.A$ and $\texttt{ct}.{\bar{B}}$   for $ A  \in \mathbb{R} ^{m \times n} $, $ B   \in \mathbb{R} ^{n \times p}$ and $B \xmapsto{\texttt{Volley Revolver Encoding}} {\bar{B}} \in \mathbb{R} ^{m \times n} $
        \Ensure The encrypted resulting matrixs $\texttt{ct}.C $  for $ C  \in \mathbb{R} ^{m \times p} $ of the matrix product $ A  \cdot B   $

        \State Set $  C  \gets \boldsymbol 0 $
        \Comment{$ C $: To accumulate intermediate matrices }
        \State $\texttt{ct}.C  \gets  \texttt{Enc}_\textit{pk}(C) $
        
        \LineComment The outer loop (could be computed in parallel)
        \For{$idx := 0$ to $p-1$}
            \State $\texttt{ct}.T \gets \texttt{RowShifter}(\texttt{ct}.{\bar{B}}, p, idx)$
            \State $\texttt{ct}.T  \gets  \texttt{Mul}(\texttt{ct}.A, \texttt{ct}.T )$
            \State $\texttt{ct}.T  \gets  \texttt{SumColVec}( \texttt{ct}.T )$  
            \LineComment Build a specifically-designed matrix to clean up the redundant values            
            \State Set $ F \gets \boldsymbol 0 $
            \Comment{$F \in \mathbb R^{m \times n}$}
            \For{$i := 1$ to $m$}
                \State $ F[i][(i+idx)\%n] \gets 1 $
            \EndFor
            \State $\texttt{ct}.T  \gets  \texttt{cMul}(F, \texttt{ct}.T)$           
            \LineComment To accumulate the intermediate results
            \State $\texttt{ct}.C  \gets  \texttt{Add}(\texttt{ct}.C, \texttt{ct}.T)$    
              
        \EndFor
          
        \State \Return $ \texttt{ct}.C $
        \end{algorithmic}
       \label{alg:Homomorphic matrix multiplication}
\end{algorithm}

The computational workflow of the proposed encoding scheme, particularly in the special case where $m = p$, admits an intuitive mechanical interpretation. Specifically, while the first matrix $\mathbf{A}$ remains fixed, the second matrix $\mathbf{B}$ undergoes successive cyclic rotations, resembling a revolver chamber that discharges multiple rounds simultaneously. This conceptual analogy motivates the name of our encoding strategy, termed \emph{\texttt{Volley Revolver}}.

In practical scenarios where $m \bmod p = 0$, the $\texttt{RowShifter}$ operation degenerates into a single rotation, namely
\[
\texttt{RowShifter} = \texttt{Rot}(\texttt{ct}, n),
\]
which is considerably more efficient than the general case. Consequently, this configuration should be preferred whenever feasible. In the context of neural network inference, such a condition can be naturally satisfied by constraining the number of neurons in each fully connected layer to be a power of two.

\subsection{Homomorphic Convolution Operation}

In this subsection, we present a homomorphic convolution algorithm under a simplifying but unrealistic assumption, and subsequently extend the idea to practical settings. We begin by describing an idealized scenario in which a single grayscale image of size $h \times w$ is perfectly packed into a single ciphertext without any unused slots, i.e., the ciphertext slot capacity $N$ coincides exactly with $h \times w$. Although such a setting is rarely achievable in practice, it serves as a useful conceptual foundation.

We then demonstrate how the same technique can be generalized to support the simultaneous convolution of multiple images of arbitrary sizes by interpreting a single ciphertext as a collection of several \emph{virtual ciphertexts}. For clarity of exposition, we restrict our discussion to grayscale images and assume that the entire dataset is encrypted into a single ciphertext.

\paragraph{An Impractical Baseline Algorithm.}
Let $I \in \mathbb{R}^{h \times w}$ denote a grayscale image and let $K \in \mathbb{R}^{k \times k}$ be a convolution kernel with bias term $k_0$, where $h > k$ and $w > k$. Under the assumption that $I$ is encrypted into a ciphertext $\texttt{ct}.I$ with neither vacant nor overflowing slots, we describe an efficient homomorphic algorithm for evaluating the convolution operation. Throughout this subsection, we assume a unit stride $(1,1)$ and adopt a valid convolution setting without padding.

Prior to executing the convolution, the kernel $K$ is processed by an auxiliary operation, denoted as $\texttt{Kernelspanner}$. This procedure generates $k^2$ ciphertexts for the common case where $h \ge 2k-1$ and $w \ge 2k-1$. Each ciphertext encrypts a matrix $P_i \in \mathbb{R}^{h \times w}$ for $1 \le i \le k^2$, obtained by embedding the kernel coefficients into the image-sized space according to a predefined spanning pattern.

As a concrete example, consider the case where $h = 4$, $w = 4$, and $k = 2$. The $\texttt{Kernelspanner}$ operation produces four ciphertexts corresponding to the kernel entries, while the kernel bias $k_0$ is used to generate an additional ciphertext:
\begin{align*}
&
\begin{bmatrix}
 k_{1} & k_{2} \\
 k_{3} & k_{4}
\end{bmatrix}
\xmapsto[\mathbb{R}^{k \times k} \mapsto k^2 \cdot \mathbb{R}^{h \times w}]{\texttt{Kernelspanner}} \\
&\quad\quad\quad
\begin{aligned}
&Enc\begin{bmatrix}
 k_{1} & k_{2} & k_{1} & k_{2} \\
 k_{3} & k_{4} & k_{3} & k_{4} \\
 k_{1} & k_{2} & k_{1} & k_{2} \\
 k_{3} & k_{4} & k_{3} & k_{4}
\end{bmatrix},
&&
Enc\begin{bmatrix}
 0 & k_{1} & k_{2} & 0 \\
 0 & k_{3} & k_{4} & 0 \\
 0 & k_{1} & k_{2} & 0 \\
 0 & k_{3} & k_{4} & 0
\end{bmatrix},
\\[1em]
&Enc\begin{bmatrix}
 0 & 0 & 0 & 0 \\
 0 & k_{1} & k_{2} & 0 \\
 0 & k_{3} & k_{4} & 0 \\
 0 & 0 & 0 & 0
\end{bmatrix},
&&
Enc
\begin{bmatrix}
 0 & 0 & 0 & 0 \\
 k_{1} & k_{2} & k_{1} & k_{2} \\
 k_{3} & k_{4} & k_{3} & k_{4} \\
 0 & 0 & 0 & 0
\end{bmatrix}.
\end{aligned}
\\[1em]
&
\begin{bmatrix}
 k_{0}
\end{bmatrix}
\mapsto
Enc
\begin{bmatrix}
 k_{0} & k_{0} & k_{0} & 0 \\
 k_{0} & k_{0} & k_{0} & 0 \\
 k_{0} & k_{0} & k_{0} & 0 \\
 0 & 0 & 0 & 0
\end{bmatrix}.
&
\end{align*}

The homomorphic convolution algorithm additionally maintains an accumulator ciphertext $\texttt{ct}.R$, which is initialized by encrypting the bias term $k_0$. The algorithm proceeds for $k^2$ iterations, where the $i$-th iteration ($1 \le i \le k^2$) consists of the following steps:

\begin{itemize}
  \item \textbf{Step 1.} Compute the element-wise product of the image ciphertext $\texttt{ct}.I$ and the kernel ciphertext $\texttt{ct}.P_i$:
  \[
  \texttt{ct}.IP_i = \texttt{Mul}(\texttt{ct}.I, \texttt{ct}.P_i).
  \]

  \item \textbf{Step 2.} Apply the procedure $\texttt{SumForConv}$ to $\texttt{ct}.IP_i$ in order to aggregate values within each $k \times k$ convolution window, yielding a ciphertext $\texttt{ct}.D$.

  \item \textbf{Step 3.} Multiply $\texttt{ct}.D$ by a carefully constructed plaintext mask to eliminate irrelevant entries, producing the filtered ciphertext $\texttt{ct}.\bar{D}$.

  \item \textbf{Step 4.} Update the accumulator via homomorphic addition:
  \[
  \texttt{ct}.R \leftarrow \texttt{Add}(\texttt{ct}.R, \texttt{ct}.\bar{D}).
  \]
\end{itemize}

Notably, Steps~1--3 can be executed independently for each $i$, enabling full parallelization across $k^2$ threads. A detailed description of the complete homomorphic convolution procedure is provided in Algorithm~\ref{alg:Homomorphic convolution operation}. The overall computational complexity of the algorithm is summarized in Table~\ref{tab3}.

\begin{table}[bht]
\centering
\caption{ Complexity and required depth of Algorithm~\ref{alg:Homomorphic convolution operation} }
\label{tab3}
\begin{tabular}{|c||c|c|c|c|c|}
\hline
$\texttt{Step}$     & $\texttt{Add}$  & $\texttt{cMult}$  & $\texttt{Rot}$  & $\texttt{Mult}$   \\
\hline\hline
$\texttt{1}$        &   0      &   0       &   0       &   1        \\
\hline
$\texttt{2}$        &   $2k$      &   1       &   $2k$       &   0        \\
\hline
$\texttt{3}$        &  0   &   1       &   0       &   0           \\
\hline
$\texttt{4}$        &  1       &   0       &   0       &   0           \\
\hline\hline
$\texttt{Total}$    &  $O(k^3)$    &   $O(k^2)$       &  $O(k^3)$      &    $O(k^2)$               \\
\hline
\end{tabular}
\end{table}

\begin{figure}[htp]
\centering
\includegraphics[scale=.4]{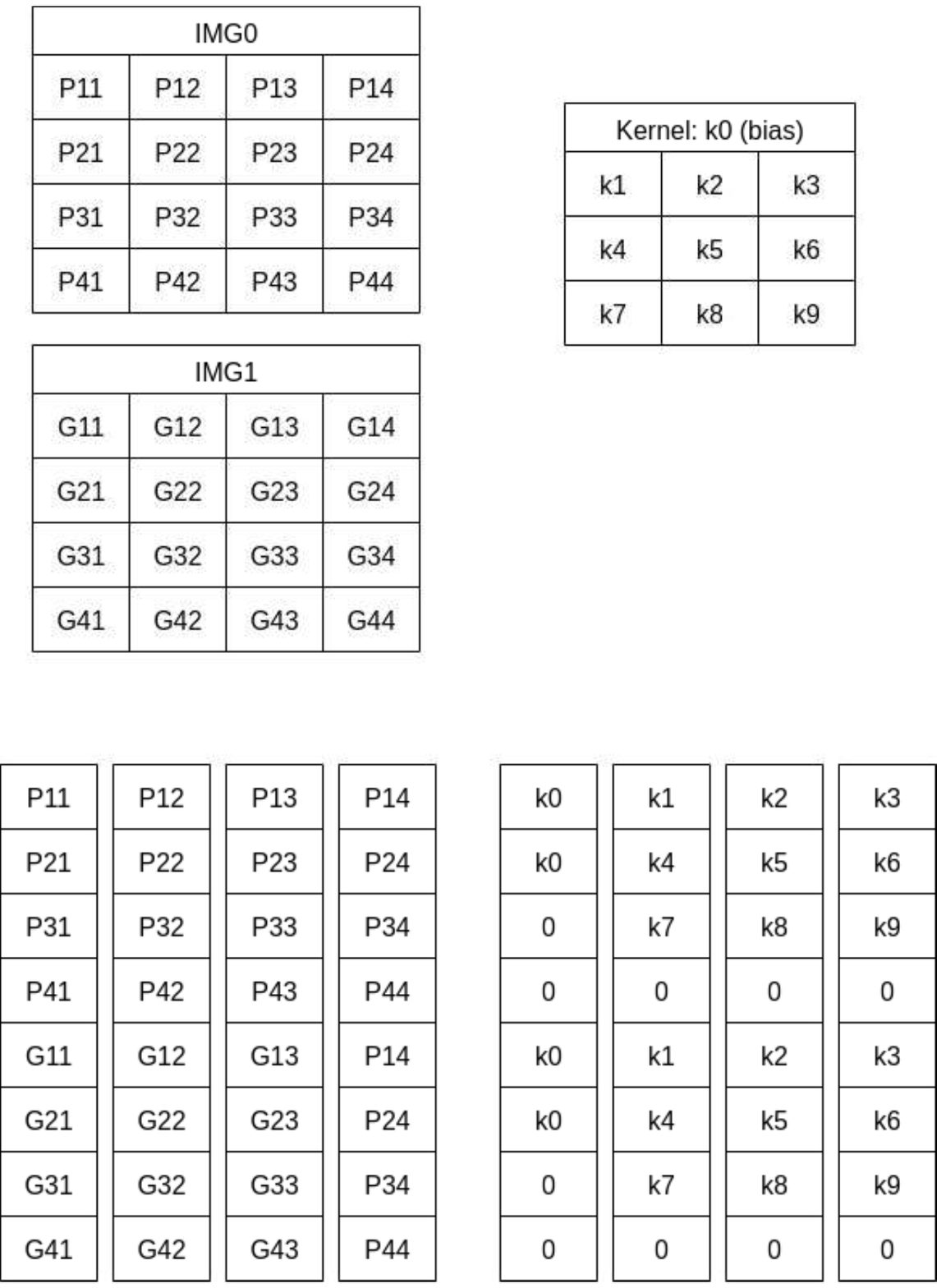}
\caption{
 Our convolution operation algorithm with $h = 3$, $w = 4$ and $k = 3$}
\label{convolution calculation}
\end{figure}

Figure~\ref{convolution calculation} illustrates a concrete example of the proposed algorithm with parameters $h = 3$, $w = 4$, and $k = 3$.

\begin{algorithm}[bh]
    \caption{Homomorphic convolution operation}
     \begin{algorithmic}[1]
        \Require An encrypted Image $ct.I$ for $I \in \mathbb{R}^{h \times w}$ and a kernel $K$ of size $k \times k$ with its bias $k_0$ 
        \Ensure The encrypted resulting image $ct.I_s $ where $I_s$ has the same size as $I$

		\LineComment The Third Party performs $\texttt{Kernelspanner}$ and prepares the ciphertext encrypting  kernel bias 
        \State  $\texttt{ct}.S_{[i]} \gets \texttt{Kernelspanner}(K, h,w) $
        \Comment{ $1 \le i \le k^2$ }
        
        \State Set $I_s \gets  \boldsymbol 0$
        \Comment{$I_s \in \mathbb R^{h \times w}$}
        \For{$i := 1$ to $h - k + 1$}
            \For{$j := 1$ to $w - k + 1$}
                \State $I_s[i][j] \gets k_0 $
            \EndFor
        \EndFor        
        \State $\texttt{ct}.I_s  \gets  \texttt{Enc}_\textit{pk}(I_s) $ 
        
        \LineComment So begins the Cloud its work
        \For{$i := 0$ to $k-1$}
            \For{$j := 0$ to $k-1$}
                \State $\texttt{ct}.T  \gets  \texttt{Mul}(\texttt{ct}.I, \texttt{ct}.S_{[i \times k + j + 1]}  )$
                \State $\texttt{ct}.T  \gets  \texttt{SumForConv}( \texttt{ct}.T )$ 
            \LineComment Design a matrix to filter out the redundant values
            \State Set $ F \gets \boldsymbol 0 $
            \Comment{$F \in \mathbb R^{m \times n}$}
                \For{$hth := 0$ to $h-1$}
                    \For{$wth := 0$ to $w-1$}
                        \If{$(wth-i) \bmod k = 0 $ {\bf and}  $wth + k \leq w$  {\bf and} \\ \hspace{2.34cm} $(hth-j) \bmod k = 0$   {\bf and} $hth + k \leq h $}

                             \State $F[hth][wth] \gets 1$

                        \EndIf
                    \EndFor
                \EndFor  
                    \State $\texttt{ct}.T  \gets  \texttt{cMul}(F, \texttt{ct}.T)$           
            \LineComment To accumulate the intermediate results
            \State $\texttt{ct}.I_s  \gets  \texttt{Add}(\texttt{ct}.I_s, \texttt{ct}.T)$   
            \EndFor
        \EndFor

        \State \Return $\texttt{ct}.I_s $
        \end{algorithmic}
       \label{alg:Homomorphic convolution operation}
\end{algorithm}

Next, we demonstrate how the aforementioned prototype homomorphic convolution algorithm can be transformed into a practical and efficient solution suitable for real-world deployments.

\subsection{Encoding Method for Convolution Operation}

For ease of presentation, we assume that the dataset
$X \in \mathbb{R}^{m \times f}$ can be encrypted into a single ciphertext
$\texttt{ct}.X$, where $m$ is a power of two. All samples are assumed to be grayscale images of identical resolution $h \times w$.
The proposed \texttt{Volley Revolver} framework encodes the dataset in matrix form using the database-style packing technique introduced in~\cite{IDASH2018Andrey}, which enables the evaluation of arbitrary convolutional neural network layers within a unified ciphertext layout.

In most practical settings, the condition $h \times w < f$ holds; in such cases, the remaining entries are padded with zeros. Moreover, \texttt{Volley Revolver} extends the original database encoding scheme by introducing additional structural operations that reinterpret the two-dimensional dataset matrix as a logical three-dimensional tensor.

Algorithm~\ref{alg:Homomorphic convolution operation} provides a feasible approach for homomorphic convolution under idealized assumptions. However, its practicality is limited by two major constraints. First, it assumes that the image size exactly matches the plaintext slot capacity, a condition that rarely holds in realistic HE parameterizations. Second, the algorithm processes only a single image per ciphertext, resulting in suboptimal throughput for real-world workloads.

To overcome these limitations, \texttt{Volley Revolver} introduces a set of simulated operations on the ciphertext $\texttt{ct}.X$ that logically reinterpret a single real ciphertext as a collection of multiple \emph{virtual ciphertexts}. Each virtual ciphertext corresponds to an individual image and supports a full suite of homomorphic operations. Collectively, these simulated operations enable the ciphertext to behave as a three-dimensional structure, where the first dimension indexes images and the remaining dimensions correspond to spatial image coordinates.

Since modern HE schemes typically offer a large number of plaintext slots, a single ciphertext can pack multiple images simultaneously. As machine learning inference applies identical computations to each sample, the same homomorphic operations can be executed in parallel across all virtual ciphertexts by operating directly on the underlying real ciphertext.

Concretely, a ciphertext encrypting the dataset
$X \in \mathbb{R}^{m \times f}$ can be viewed as simulating $m$ virtual ciphertexts $\texttt{vct}_i$ for $1 \le i \le m$, as illustrated below. Alternatively, each virtual ciphertext may be interpreted as encrypting a two-dimensional image matrix:
 
\begin{align*}
& Enc
\begin{bmatrix}
 I_{[1][1]}^{(1)} & I_{[1][2]}^{(1)} & \ldots & I_{[h][w]}^{(1)} & 0 & \ldots & 0 \\
 I_{[1][1]}^{(2)} & I_{[1][2]}^{(2)} & \ldots & I_{[h][w]}^{(2)} & 0 & \ldots & 0 \\
 \vdots & \vdots & \ddots & \vdots & \vdots & \ddots & \vdots \\
 I_{[1][1]}^{(m)} & I_{[1][2]}^{(m)} & \ldots & I_{[h][w]}^{(m)} & 0 & \ldots & 0
\end{bmatrix}
\longrightarrow
\\[0.8em]
&
\begin{bmatrix}
 \texttt{vEnc}\!
 \begin{bmatrix}
  I_{[1][1]}^{(1)} & I_{[1][2]}^{(1)} & \ldots & I_{[h][w]}^{(1)}
 \end{bmatrix}\!
  & 0 & \ldots & 0 \\[0.4em]
 \texttt{vEnc}\!
 \begin{bmatrix}
  I_{[1][1]}^{(2)} & I_{[1][2]}^{(2)} & \ldots & I_{[h][w]}^{(2)}
 \end{bmatrix}\!
  & 0 & \ldots & 0 \\
 \vdots & \vdots & \ddots & \vdots \\[0.4em]
 \texttt{vEnc}\!
 \begin{bmatrix}
  I_{[1][1]}^{(m)} & I_{[1][2]}^{(m)} & \ldots & I_{[h][w]}^{(m)}
 \end{bmatrix}\!
  & 0 & \ldots & 0
\end{bmatrix},
\\[1em]
&\text{or}\\[1em]
& Enc
\begin{bmatrix}
 I_{[1][1]}^{(1)} & I_{[1][2]}^{(1)} & \ldots & I_{[h][w]}^{(1)} & 0 & \ldots & 0 \\
 I_{[1][1]}^{(2)} & I_{[1][2]}^{(2)} & \ldots & I_{[h][w]}^{(2)} & 0 & \ldots & 0 \\
 \vdots & \vdots & \ddots & \vdots & \vdots & \ddots & \vdots \\
 I_{[1][1]}^{(m)} & I_{[1][2]}^{(m)} & \ldots & I_{[h][w]}^{(m)} & 0 & \ldots & 0
\end{bmatrix}
\longrightarrow
\\[0.8em]
& Enc
\begin{bmatrix}
 \texttt{vEnc}\!
 \begin{bmatrix}
  I_{[1][1]}^{(1)} & \ldots & I_{[1][w]}^{(1)} \\
  \vdots & \ddots & \vdots \\
  I_{[h][1]}^{(1)} & \ldots & I_{[h][w]}^{(1)}
 \end{bmatrix}\!
  & 0 & \ldots & 0 \\[0.4em]
 \vdots & \vdots & \ddots & \vdots \\[0.4em]
 \texttt{vEnc}\!
 \begin{bmatrix}
  I_{[1][1]}^{(m)} & \ldots & I_{[1][w]}^{(m)} \\
  \vdots & \ddots & \vdots \\
  I_{[h][1]}^{(m)} & \ldots & I_{[h][w]}^{(m)}
 \end{bmatrix}\!
  & 0 & \ldots & 0
\end{bmatrix}.
\end{align*}

Analogous to a standard HE ciphertext, a virtual ciphertext supports a collection of virtual homomorphic operations, including
$\texttt{vEnc}$, $\texttt{vDec}$, $\texttt{vAdd}$, $\texttt{vMul}$,
$\texttt{vRescale}$, $\texttt{vBootstrapping}$, and $\texttt{vRot}$.
With the exception of $\texttt{vRot}$, all virtual operations are inherited directly from their corresponding HE operations. In particular, the homomorphic operations
$\texttt{Add}$, $\texttt{Mul}$, $\texttt{Rescale}$, and $\texttt{Bootstrapping}$
naturally induce the virtual operations
$\texttt{vAdd}$, $\texttt{vMul}$, $\texttt{vRescale}$, and
$\texttt{vBootstrapping}$.

The virtual rotation operation $\texttt{vRot}$ is unique in that it requires two coordinated rotations on the underlying real ciphertext, as detailed in Algorithm~\ref{alg:vRot}. In practice, only virtual rotations are required to emulate convolution behavior. Specifically, the operation
$\texttt{vRot}(\texttt{ct}, r)$ rotates each virtual ciphertext within the real ciphertext $\texttt{ct}$ leftward by $r$ positions, yielding the following transformation:

 \begin{align*} 
 &
 \begin{bmatrix}
 	\texttt{vEnc}
 	\begin{mbmatrix}
  	I_{[1][1]}^{(1)}  &   \ldots  &  I_{[r/w][r\%w]}^{(1)}  & 	I_{[(r+1)/w][(r+1)\%w]}^{(1)}  &  \ldots   &  I_{[h][w]}^{(1)}    
 	\end{mbmatrix}        \\
 	\vdots                \\
	\texttt{vEnc}
 	\begin{mbmatrix}
	I_{[1][1]}^{(m)}  &   \ldots  &  I_{[r/w][r\%w]}^{(m)}  & 	I_{[(r+1)/w][(r+1)\%w]}^{(m)}  &  \ldots   &  I_{[h][w]}^{(m)}   
 	\end{mbmatrix}        \\
 \end{bmatrix}  \\
 &   \hspace{5.02cm} \downarrow{\texttt{vRot}(\texttt{ct}, r)}   \\ 
 &
 \begin{bmatrix}
 	\texttt{vEnc}
 	\begin{mbmatrix}
  	I_{[(r+1)/w][(r+1)\%w]}^{(1)}  &  \ldots   &  I_{[h][w]}^{(1)}      &     I_{[1][1]}^{(1)}  &   \ldots  &  I_{[r/w][r\%w]}^{(1)}     
 	\end{mbmatrix}        \\
 	\vdots                \\
	\texttt{vEnc}
 	\begin{mbmatrix}
	I_{[(r+1)/w][(r+1)\%w]}^{(m)}  &  \ldots   &  I_{[h][w]}^{(m)}      &     I_{[1][1]}^{(1)}  &   \ldots  &  I_{[r/w][r\%w]}^{(m)}     
 	\end{mbmatrix}       \\
 \end{bmatrix}.  
\end{align*}


Given two collections of virtual ciphertexts
$\{\texttt{vct}_{[i]}\}$ and $\{\texttt{vct}_{[j]}\}$ residing in ciphertexts
$\texttt{ct}_1$ and $\texttt{ct}_2$, respectively, the virtual multiplication
$\texttt{vMul}(\texttt{vct}_{[i]}, \texttt{vct}_{[j]})$ is realized by a single homomorphic multiplication on the real ciphertexts and produces element-wise products across all virtual ciphertexts. 
\begin{align*} 
& Enc
\begin{bmatrix}
 \texttt{vEnc}\!\begin{bmatrix} I_{[1][1]}^{(1)} & \ldots & I_{[h][w]}^{(1)} \end{bmatrix} & 0 & \ldots & 0 \\
 \vdots & \vdots & \ddots & \vdots \\
 \texttt{vEnc}\!\begin{bmatrix} I_{[1][1]}^{(m)} & \ldots & I_{[h][w]}^{(m)} \end{bmatrix} & 0 & \ldots & 0
\end{bmatrix}  
\\[0.8em]
&\hspace{2.02cm} \otimes
\\[0.8em]
& Enc
\begin{bmatrix}
 \texttt{vEnc}\!\begin{bmatrix} G_{[1][1]}^{(1)} & \ldots & G_{[h][w]}^{(1)} \end{bmatrix} & 0 & \ldots & 0 \\
 \vdots & \vdots & \ddots & \vdots \\
 \texttt{vEnc}\!\begin{bmatrix} G_{[1][1]}^{(m)} & \ldots & G_{[h][w]}^{(m)} \end{bmatrix} & 0 & \ldots & 0
\end{bmatrix}  
\\[0.8em]
& \hspace{2.02cm} \Downarrow{\texttt{vMul}(\texttt{vct}_{[i]}, \texttt{vct}_{[j]})}
\\[0.8em]
& Enc
\begin{bmatrix}
 \texttt{vEnc}\!\begin{bmatrix} I_{[1][1]}^{(1)} \cdot G_{[1][1]}^{(1)} & \ldots & I_{[h][w]}^{(1)} \cdot G_{[h][w]}^{(1)} \end{bmatrix} & \ldots  \\
 \vdots & \vdots  \\
 \texttt{vEnc}\!\begin{bmatrix} I_{[1][1]}^{(m)} \cdot G_{[1][1]}^{(m)} & \ldots & I_{[h][w]}^{(m)} \cdot G_{[h][w]}^{(m)} \end{bmatrix} & \ldots 
\end{bmatrix}.
\end{align*}

The virtual addition operation $\texttt{vAdd}$ follows an analogous pattern and yields element-wise summation across corresponding virtual slots.

 \begin{align*} 
& \hspace{0.22cm} Enc
\begin{bmatrix}
 \texttt{vEnc}\begin{bmatrix}
  I_{[1][1]}^{(1)}  & \ldots &  I_{[1][w]}^{(1)} \\
  \vdots            & \ddots &  \vdots           \\
  I_{[h][1]}^{(1)}  & \ldots &  I_{[h][w]}^{(1)}  
 \end{bmatrix} & 0 & \ldots & 0 \\
 \vdots & \vdots & \ddots & \vdots \\
 \texttt{vEnc}\begin{bmatrix}
  I_{[1][1]}^{(m)}  & \ldots &  I_{[1][w]}^{(m)} \\
  \vdots            & \ddots &  \vdots           \\
  I_{[h][1]}^{(m)}  & \ldots &  I_{[h][w]}^{(m)}  
 \end{bmatrix} & 0 & \ldots & 0
\end{bmatrix}  
\\[0.8em]
& \hspace{3.02cm} \oplus
\\[0.8em]
& \hspace{0.22cm} Enc
\begin{bmatrix}
 \texttt{vEnc}\begin{bmatrix}
  G_{[1][1]}^{(1)}  & \ldots &  G_{[1][w]}^{(1)} \\
  \vdots            & \ddots &  \vdots           \\
  G_{[h][1]}^{(1)}  & \ldots &  G_{[h][w]}^{(1)}  
 \end{bmatrix} & 0 & \ldots & 0 \\
 \vdots & \vdots & \ddots & \vdots \\
 \texttt{vEnc}\begin{bmatrix}
  G_{[1][1]}^{(m)}  & \ldots &  G_{[1][w]}^{(m)} \\
  \vdots            & \ddots &  \vdots           \\
  G_{[h][1]}^{(m)}  & \ldots &  G_{[h][w]}^{(m)}  
 \end{bmatrix} & 0 & \ldots & 0
\end{bmatrix}  
\\[0.8em]
& \hspace{3.02cm}\Downarrow{\texttt{vAdd}(\texttt{vct}_{[i]}, \texttt{vct}_{[j]})}
\\[0.8em]
& \hspace{0.02cm} Enc
\begin{bmatrix}
 \texttt{vEnc}\begin{bmatrix}
  I_{[1][1]}^{(1)} + G_{[1][1]}^{(1)}  & \ldots &  I_{[1][w]}^{(1)} + G_{[1][w]}^{(1)} \\
  \vdots            & \ddots &  \vdots                                   \\
  I_{[h][1]}^{(1)} + G_{[h][1]}^{(1)}  & \ldots &  I_{[h][w]}^{(1)} + G_{[h][w]}^{(1)}  
 \end{bmatrix} &  \ldots \\
 \vdots & \vdots \\
 \texttt{vEnc}\begin{bmatrix}
  I_{[1][1]}^{(m)} + G_{[1][1]}^{(m)}  & \ldots &  I_{[1][w]}^{(m)} + G_{[1][w]}^{(m)} \\
  \vdots            & \ddots &  \vdots                                   \\
  I_{[h][1]}^{(m)} + G_{[h][1]}^{(m)}  & \ldots &  I_{[h][w]}^{(m)} + G_{[h][w]}^{(m)}  
 \end{bmatrix} &  \ldots 
\end{bmatrix}.
\end{align*}


\begin{algorithm}[htp]
    \caption{vRot: virtual rotation operations on a set of virtual ciphertexts}
     \begin{algorithmic}[1]
        \Require a ciphertext $\texttt{ct}.X$ encrypting a  dataset matrix $X$ of size $m \times n$, each row of which encrypts an image of size $ h \times w$ such that $ h \times w \le n$, and the number $r$ of rotations to the left 
        \Ensure a ciphertext $\texttt{ct}.R$ encrypting the resulting  matrix $R$ of the same size as $X$

        \State Set $ R \gets \boldsymbol 0 $
        \Comment{$R \in \mathbb R^{m \times n}$}
        \State $\texttt{ct}.R  \gets  \texttt{Enc}_\textit{pk}(R) $
        
        \LineCommentStep1 Rotate the real ciphertext to the left by $r$ positions first and then clean up  the garbages values
        \State $\texttt{ct}.T \gets \texttt{Rot}(\texttt{ct}.X, r)$
        \LineComment Build a specially designed matrix
        \State Set $ F_1 \gets \boldsymbol 0 $
        \Comment{$F_1 \in \mathbb R^{m \times n}$}
        \For{$i := 1$ to $m$}
            \For{$j := 1$ to $h \times w - r$}
                \State $ F_1[i][j] \gets 1 $
            \EndFor
        \EndFor
        \State $\texttt{ct}.T_1  \gets  \texttt{cMul}(F_1, \texttt{ct}.T)$
          
        \LineCommentStep2 Rotate the real ciphertext to the right by $h \times w - r$ positions (the same as to the left by $m \times n - h \times w + r$ positions) first and then clean up the garbage values
        \State $\texttt{ct}.P \gets \texttt{Rot}(\texttt{ct}.X, m \times n - h \times w + r )$
        \LineComment Build a specilal designed matrix 
        \State Set $ F_2 \gets \boldsymbol 0 $
        \Comment{$F_2 \in \mathbb R^{m \times n}$}
        \For{$i := 1$ to $m$}
            \For{$j := h \times w - r + 1$ to $h \times w$}
                \State $ F_2[m][j] \gets 1 $
            \EndFor
        \EndFor
        \State $\texttt{ct}.T_2  \gets  \texttt{cMul}(F_2, \texttt{ct}.P)$
          
        \LineComment  concatenate            
        \State $\texttt{ct}.R  \gets  \texttt{Add}(\texttt{ct}.T_1, \texttt{ct}.T_2)$
 
        \State \Return $ \texttt{ct}.R $
        \end{algorithmic}
       \label{alg:vRot}
\end{algorithm}

By integrating all the components described above, Algorithm~\ref{alg:Homomorphic convolution operation} can be leveraged to perform convolution operations over multiple images concurrently by operating on simulated virtual ciphertexts. A key advantage of this simulation framework is that any sequence of homomorphic operations applied to a real ciphertext implicitly induces the corresponding operations on all virtual ciphertexts embedded within it. This property enables efficient batch processing and suffices for real-world inference workloads.

\section{Secure Inference}

In practical deployments, datasets are often too large to be encrypted into a single ciphertext due to plaintext slot limitations. To address this issue, the dataset can be partitioned into multiple mini-batches, each containing an equal number of samples. Each mini-batch is then encrypted into an independent ciphertext. This strategy preserves the applicability of the proposed algorithms while naturally enabling higher levels of parallelism.

\subsection{Parallelized Matrix Multiplication}

Algorithm~\ref{alg:Homomorphic matrix multiplication} is originally formulated under the assumption that all data can be packed into a single ciphertext. Nevertheless, its performance can be substantially improved by exploiting hardware-level parallelism. Consider two matrices $A \in \mathbb{R}^{m \times n}$ and $B \in \mathbb{R}^{n \times p}$, where the algorithm requires $O(p \log n)$ ciphertext rotations. If $p$ hardware threads are available, the computation can be parallelized across columns of $B$, reducing the number of rotations per thread to $O(\log p)$.

When the dataset exceeds the capacity of a single ciphertext, matrices $A$ and $B$ can be partitioned into sets of ciphertexts $\{A_1, \ldots, A_i\}$ and $\{B_1, \ldots, B_j\}$. Under this setting, the same algorithm remains applicable. Each ciphertext in $\{A_\ell\}$ can be multiplied with each ciphertext in $\{B_r\}$ independently, allowing matrix multiplication to be parallelized at the ciphertext level. From a mathematical perspective, this approach corresponds to computing products over submatrices and preserves correctness by linearity.

Therefore, even when multiple ciphertexts are involved, the proposed method continues to support efficient and scalable parallel computation by distributing workloads across ciphertext pairs.

\subsection{Parallelized Convolution Operation}

By exploiting the simulated operations on real ciphertexts, Algorithm~\ref{alg:Homomorphic convolution operation} can be used to perform convolution over multiple images simultaneously via virtual ciphertexts. The effective time complexity can be further reduced through amortization, depending on how many virtual ciphertexts are embedded in a single real ciphertext. Specifically, if a ciphertext simulates $m$ virtual ciphertexts, then convolution can be performed on $m$ images concurrently, reducing the amortized computation cost to $\frac{1}{m}$ of the baseline.

Recall that the convolution kernel $K$ is preprocessed using $\texttt{Kernelspanner}$ to generate $k^2$ ciphertexts $K_1, \ldots, K_{k^2}$, each encrypting an $h \times w$ matrix. Given a ciphertext $\texttt{ct}.I$ encrypting a single-channel image of size $h \times w_I$, the partial convolution operations between $\texttt{ct}.I$ and the kernel ciphertexts $K_i$ can be executed independently. Moreover, if $k^2$ hardware threads are available, these operations can be fully parallelized, significantly accelerating the convolution process.

For color images or intermediate feature maps with multiple channels, each channel is encrypted into a separate ciphertext and processed independently using Algorithm~\ref{alg:Homomorphic convolution operation}. In typical convolutional layers, each kernel contains the same number of channels as the input image. Consequently, ciphertexts encrypting different input channels are multiplied with the corresponding kernel-channel ciphertexts, producing partial convolution results. These intermediate ciphertexts are then accumulated via homomorphic addition to obtain the final convolution output. Convolutions associated with different kernels can also be computed in parallel, yielding substantial performance gains through multi-level parallelism across channels, kernels, and ciphertexts.

The proposed \texttt{Volley Revolver} framework supports the construction of convolutional neural networks of arbitrary depth. However, as network depth increases, the multiplicative depth grows accordingly, necessitating the use of bootstrapping to refresh ciphertexts. While bootstrapping enables unbounded-depth computation, it introduces additional computational overhead and remains a dominant cost factor.

\subsection{Homomorphic CNN Evaluation}

Following a convolutional layer, it is often necessary to reshape the encrypted data structure, as the convolution operation performed by $\texttt{SumForConv}$ reduces the spatial dimensions of the image. Consequently, the resulting ciphertext encoding may no longer conform to the original \texttt{Volley Revolver} layout. An illustrative example is shown below:
 \begin{align*} 
 & Enc
 \begin{bmatrix}
  \texttt{vEnc}
 \begin{mbmatrix}
  I_{[1][1]}^{(1)}    &   I_{[1][2]}^{(1)}    &  \ldots &  I_{[3][3]}^{(1)}  
 \end{mbmatrix}     &   0   &  \ldots  & 0   \\
  \vdots    &   \vdots   &  \ddots  & \vdots   \\
 \end{bmatrix} \\ 
 & = Enc
 \begin{bmatrix}
 \texttt{vEnc}
 \begin{mbmatrix}
  I_{[1][1]}^{(1)}    &   I_{[1][2]}^{(1)}    &   I_{[1][3]}^{(1)}  \\
  I_{[2][1]}^{(1)}    &  I_{[2][2]}^{(1)}     &    I_{[2][3]}^{(1)}            \\
  I_{[3][1]}^{(1)}    &  I_{[3][2]}^{(1)}     &  I_{[3][3]}^{(1)}    \\  
 \end{mbmatrix}     &   0   &  \ldots  & 0   \\
  \vdots    &   \vdots   &  \ddots  & \vdots   \\
 \end{bmatrix} 
 \\
 &   \hspace{1.65cm} \big\Downarrow{\texttt{SumForConv}(\cdot,2,2) }  
 \\ 
  &  Enc
 \begin{bmatrix}
 \texttt{vEnc}
 \begin{mbmatrix}
  J_{[1][1]}^{(1)}    &   J_{[1][2]}^{(1)}    &   0  \\
  J_{[2][1]}^{(1)}    &  J_{[2][2]}^{(1)}     &    0            \\
  0    &  0    &  0   \\  
 \end{mbmatrix}     &   0   &  \ldots  & 0   \\
  \vdots    &   \vdots   &  \ddots  & \vdots   \\
 \end{bmatrix} 
 \\
  &   \hspace{1.65cm} \big\Downarrow{\texttt{ReForm}(\cdot,2,2) }  
 \\
  & \ \ \ \ \ Enc
 \begin{bmatrix}
 \texttt{vEnc}
 \begin{mbmatrix}
  J_{[1][1]}^{(1)}    &   J_{[1][2]}^{(1)}      \\
  J_{[2][1]}^{(1)}    &  J_{[2][2]}^{(1)}               \\
 \end{mbmatrix}  & 0   &  \ldots     &   0   &   0   &  \ldots  & 0   \\
  \vdots     & \vdots   &  \ldots     &   \vdots   &   \vdots   &  \ddots  & \vdots \\
 \end{bmatrix}
 \\
  & =Enc
 \begin{bmatrix}
 \texttt{vEnc}
 \begin{mbmatrix}
  J_{[1][1]}^{(1)}    &   J_{[1][2]}^{(1)}   & J_{[2][1]}^{(1)}    &  J_{[2][2]}^{(1)}               \\
 \end{mbmatrix}  & 0   &  \ldots     &   0     \\
  \vdots     & \vdots   &  \ldots     &   \vdots    \\
 \end{bmatrix}
\end{align*} 

For stride $1$, this reshaping procedure incurs $O(h)$ ciphertext rotations, $O(h)$ constant multiplications, and $O(h)$ additions, where $h$ denotes the height of the original image.

Pooling layers similarly reduce spatial dimensions and require additional reshaping operations. Since the baseline work~\cite{jiang2018secure} does not incorporate pooling layers, we omit them as well for consistency.

\paragraph*{Threat Model}
We assume that the underlying homomorphic encryption scheme satisfies IND-CPA security, meaning that ciphertexts encrypting any two messages are computationally indistinguishable. Under this assumption, all computations performed by the cloud server operate exclusively on encrypted data. As a result, a semi-honest cloud server learns no information about either the input data or the inference results, ensuring data confidentiality throughout the computation.

\paragraph*{Usage Model}
Our approach supports multiple deployment scenarios similar to those discussed in~\cite{han2018efficient}. In all cases, both the privacy-sensitive data and the CNN model parameters are encrypted before being outsourced to the cloud. A typical usage model involves three roles: the data owner, the model provider, and the cloud server. In certain scenarios, the first two roles may be assumed by the same entity.

Under this model, the responsibilities of each role are summarized as follows:
\begin{enumerate}
 \item \textbf{Data Owner.} The data owner prepares the dataset by performing preprocessing steps such as normalization and, if necessary, partitioning the dataset into mini-batches. Each mini-batch is encrypted using the database encoding method of~\cite{IDASH2018Andrey} and uploaded to the cloud.

 \item \textbf{Model Provider.} The model provider generates kernel ciphertexts using $\texttt{Kernelspanner}$ and encodes the fully connected layer weights using the \texttt{Volley Revolver} method before encryption. Polynomial approximations of activation functions (e.g., ReLU) are treated as public parameters and can be transmitted to the cloud without encryption.

 \item \textbf{Cloud Server.} The cloud server deploys the homomorphic CNN inference application and maintains the HE runtime environment. Public keys and encrypted model parameters are provisioned collaboratively with the model provider prior to inference.
\end{enumerate}

\section{Implementation}

We implement our homomorphic CNN inference framework in \texttt{C++}. 
The complete source code is publicly available at
\url{https://github.com/petitioner/HE.CNNinference}.

\subsection{Neural Network Architecture}

We adopt the same CNN architecture proposed in~\cite{jiang2018secure}, with several modifications to the hyperparameters to optimize the performance of our approach. 
The detailed configuration of the CNN used for the MNIST dataset is summarized in Table~\ref{tab1}.

\begin{table*}[htbp]
\centering
\caption{Description of our CNN on the MNIST dataset }
\label{tab1}
\begin{tabular}{|l||c|}
\hline
Layer & \multicolumn{1}{c|}{Description}  \\
\hline\hline
CONV               & 32 input images of size $28 \times 28$,  
                            4 kernels of size $3 \times 3$,     
                            stride size of ($1$, $1$)              \\
\hline
ACT-$1$     &  \mysplit{$x \mapsto -0.00015120704 + 0.4610149 \cdot x + 2.0225089 \cdot x^2  -1.4511951 \cdot x^3$}  \\
\hline
FC-$1$      & \mysplit{Fully connected layer with $26 \times 26 \times 4 = 2704$   inputs and $64$ outputs }   \\
\hline
ACT-$2$              &  \mysplit{     $x \mapsto  -1.5650465  -0.9943767 \cdot x  + 1.6794522 \cdot x^2 + 0.5350255 \cdot x^3 $ }    \\
\hline
FC-$2$            &  Fully connected layer with 64 inputs and 10 outputs   \\
\hline
\end{tabular}
\end{table*}

\subsection{Building a Model in the Clear}

Homomorphic encryption does not support the direct evaluation of non-polynomial functions such as the \texttt{ReLU} activation. 
To address this limitation, we approximate the \texttt{ReLU} function using a degree-three polynomial obtained via the least-squares method implemented in \texttt{Octave}. 
This polynomial approximation is used to initialize all activation layers, and the training process further refines the coefficients for each layer.

To construct a homomorphic-compatible model, we follow a standard training pipeline in the plaintext domain, with the only modification being the replacement of all \texttt{ReLU} activations with their polynomial approximations. 
Specifically, we train a CNN with the architecture shown in Table~\ref{tab1} on the MNIST training dataset, with input pixel values normalized to the range $[0,1]$. 
The model is implemented using the \texttt{Keras} library with a \texttt{TensorFlow} backend, which provides flexible support for custom activation functions.

After convergence, the trained model parameters are exported to CSV files for subsequent use in the encrypted domain. 
In the second stage, we reimplement the CNN using the \texttt{HEAAN} library by importing the pre-trained weights and applying HE-compatible arithmetic. 
During inference, the input images are normalized by dividing each pixel value by 255, consistent with the preprocessing used during plaintext training.

\subsection{Homomorphic CNN Inference}

\paragraph{Parameters.}
Following the notation in~\cite{IDASH2018Andrey}, we configure the homomorphic encryption parameters as follows.
The scaling factors are set to $\Delta = 2^{45}$ and $\Delta_c = 2^{20}$, and the number of slots is set to $\texttt{slots} = 32768$.
The ciphertext modulus and polynomial modulus degree are configured as $\texttt{logQ} = 1200$ and $\texttt{logN} = 16$, respectively, providing an estimated security level of approximately 80 bits.
Further details regarding parameter selection can be found in~\cite{han2018efficient, jiang2018secure}.

\subsection{Performance Evaluation}

We evaluate the performance of our implementation on the MNIST testing dataset consisting of 10,000 images.
Each image has size $28 \times 28$, with 256-level grayscale pixel values and labels ranging from zero to nine.

Under the chosen parameter configuration, a single ciphertext can pack and process 32 MNIST images simultaneously.
Accordingly, we partition the test dataset into 313 batches, padding the final batch with zeros to form a complete block.
We then execute the homomorphic CNN inference on these 313 ciphertexts.

Our method achieves a final classification accuracy of $98.61\%$ on the MNIST test set.
Each ciphertext produces 32 classification results, one for each encrypted image, by selecting the digit with the highest predicted probability.
The processing of a single ciphertext requires approximately 287 seconds on a cloud server equipped with 40 vCPUs.

In terms of communication overhead, the data owner uploads only a single ciphertext of approximately 19.8 MB to encrypt 32 input images.
The model provider transmits 52 ciphertexts, totaling approximately 1 GB, which contain the encrypted weights of the trained CNN model.

\section{Conclusion}

In this work, we proposed \texttt{Volley Revolver}, an encoding method specifically tailored for privacy-preserving neural networks under fully homomorphic encryption.
Beyond inference, this encoding strategy shows strong potential for assisting FHE-based neural network training, particularly in the backpropagation of fully connected layers, where one matrix can be revolved while the other remains fixed.

By appropriately distributing responsibilities among the data owner, model provider, and cloud server, we enable efficient homomorphic CNN inference without revealing sensitive information.
This division of labor is well suited to real-world privacy-preserving inference scenarios.

Although our experiments focus on grayscale images that can be packed into a single ciphertext, the proposed method naturally extends to large-scale color image datasets and different stride sizes using the same padding and encoding techniques.
Ongoing research indicates that our approach can be further adapted to practical applications involving high-resolution images and deeper CNN architectures.

\bibliography{HE_CNNinfer}

\end{document}